\documentclass[10pt,flushrt,tighten,preprint]{aastex}

\newcommand{\yr}{\mbox{$\rm\,yr$}}
\newcommand{\pc}{\mbox{$\rm\,pc$}}
\newcommand{\au}{\mbox{$\rm\,AU$}}
\newcommand{\msun}{\mbox{$\,M_\odot$}}
\newcommand{\kms}{\mbox{${\rm\,km\,s}^{-1}$}}
\newcommand{\beq}{\begin{equation}}
\newcommand{\eeq}{\end{equation}}
\usepackage{epsf}

\slugcomment{\sl Astronomical Journal, in press}

\shorttitle{Eccentric Fluid Disks}
\shortauthors{Statler}

\begin{document}

\title{A Simple Family of Models for Eccentric Keplerian Fluid Disks}

\author{Thomas S. Statler}
\affil{Department of Physics and Astronomy, 251B Clippinger Research
Laboratories, Ohio University, Athens, OH 45701, USA}
\email{statler@ohio.edu}

\begin{abstract}
In order to be in a long-lived configuration, the density in a fluid disk
should be constant along streamlines to prevent compressional ($PdV$) work
from being done cyclically around every orbit. In a pure Kepler potential,
flow along aligned, elliptical streamlines of constant eccentricity will
satisfy this condition. For most density profiles, differential precession
driven by the pressure gradient will destroy the alignment; however, in the
razor-thin approximation there is a family of simple equilibria in which
the precession frequency is the same at all radii. These disks may
therefore be long-lived at significant eccentricities. The density can be
made axisymmetric as $r\to 0$, while maintaining the precession
rate, by relaxing the requirement of constancy along streamlines in an
arbitrarily small transition region near the center. In the limit of small
eccentricity, the models can be seen as acoustically perturbed axisymmetric
disks, and the precession rate is shown to agree with linear theory. The
perturbation is a traveling wave similar to an ocean wave, with the fluid
rising and falling epicyclically in the gravitational field of the central
mass. The expected emission line profiles from the eccentric disks are
shown to be strongly asymmetric in general, and, in extreme cases, prone to
misinterpretation as single narrow lines with significant velocity offsets.
\end{abstract}

\keywords{galaxies: kinematics and dynamics---galaxies: nuclei---galaxies:
structure--- stars: formation---stars: novae, cataclysmic variables}

\section{Introduction}

Accretion disks, protostellar disks, and galactic disks are
almost always assumed to be axisymmetric or nearly so. This assumption is
not without justification. Disks support a rich spectrum of stable and
unstable modes which can efficiently redistribute angular momentum
while viscous, collisional, and radiative processes dissipate energy.
Circularization is then a naturally expected result.
Nonetheless, disks of finite eccentricity may play an important role in
certain astrophysical systems, and, indeed, appear to be required by
observed phenomena on scales from AU to parsecs.

Two examples illustrate this need particularly well.
At the stellar scale, eccentric disks are well established as the source of
the superhump phenomenon in short-period cataclysmic variables. The period
of these transient luminosity variations differs by a few percent from the
binary orbital period, and the difference is readily explained by a
precession of the disk \citep{Vog82,Osa85}. The eccentricity is ``pumped''
by the binary tidal field; numerical and analytic studies
\citep{Whi88,Lub91} indicate that the eccentricity can be excited at the
3:1 Lindblad resonance (but see \cite{Hee94} and \citet{Ste99} for
dissenting views).  While the eccentricity may vary between outbursts, the
repeatability of superhump light curves within an outburst \citep{Pat98}
indicates that the eccentricity is not washed away on short time scales by
differential precession.  Instead, the disk needs to precess coherently for
at least several orbital periods. The primary driver for this precession is
most likely the gravity of the secondary star, but the disk's own pressure
gradient may also be a significant influence \citep{Mur00}.

At the parsec scale, the best studied eccentric disk candidate is the
``double nucleus'' of M31 \citep{Lau93,Tre95,KoB99}.
This disk is stellar rather than gaseous, and is essentially isolated in
the Keplerian potential of the central black hole, since neither the
bulge nor the main disk has significant influence at this scale.
Assuming the disk is not a transient, its structure must
be such that its own self gravity is able to drive a coherent precession.
This condition imposes a characteristic non-monotonic radial eccentricity
profile on the disk \citep{Sta99,Sal01}, which is reproduced by dynamical
simulations \citep{Bac01}. However, whether the eccentricity can be
self-excited or must be tidally driven remains at issue \citep{Bac01,Tre01},
and the dynamical stability of such configurations is undetermined.

This paper takes up a question complementary to both of the above examples,
namely, whether isolated Keplerian fluid disks can exist in long-lived
eccentric equilibria. I will not address directly issues of hydrodynamic
stability, which are extremely complex---even in axisymmetric systems---and
far beyond the simple arguments presented here. But some necessary
constraints on disk structure can be gleaned without a lengthy stability
analysis. A slowly evolving disk should have a density that is
approximately constant around streamlines. Were this not the case, $PdV$
work would be done cyclically, and most likely irreversibly, around every
orbit, dissipating the eccentric motions. In addition, the density and
pressure must be such that perturbed streamlines can precess coherently.
In the approximation of a razor-thin two dimensional disk with a polytropic
equation of state and negligible mass, these minimal criteria for
astrophysical realism are sufficient to define a remarkably simple family
of eccentric disk models.

The arguments of the paper are laid out as follows: Section \ref{s.models}
presents the basic properties of the models. Simple scaling arguments are
used in \S\ \ref{s.scaling} to derive the radial density profile for an
assumed equation of state. Section \ref{s.precession} calculates the
pressure-driven precession, and demonstrates that in the frame that rotates
at the precession frequency, the motion of fluid elements is Keplerian to
first order in the pressure. In \S\ \ref{s.center} the slightly sticky
question of the inner boundary condition is considered. It is shown how
the central eccentricity can be taken smoothly to zero by altering the density
profile in an arbitrarily small transition region.
The structure of the models in the low-eccentricity limit is shown in
\S\ \ref{s.axisymmetric} to correspond to that of axisymmetric disks
perturbed by an $m=1$ traveling wave. The expected emission-line profiles
from unresolved disks are computed in \S\ \ref{s.lineprofiles}. Finally,
\S\ \ref{s.discussion} discusses the connection with other work on eccentric
disks, describes the possible instabilities that could affect the models,
and reiterates the main results.

\section{Disk Models\label{s.models}}

\subsection{Scaling Arguments and Radial Profile\label{s.scaling}}

The disks of this paper are assumed to be planar and razor-thin, with zero
thickness perpendicular to the plane. The fluid has zero viscosity, and is
characterized by a
surface density $\Sigma$ and a two-dimensional pressure $\Pi$, related 
by a 2-D analogue of a polytropic equation of state,
\beq\label{e.eos}
\Pi = K \Sigma^{1+1/n},
\eeq
in which $n$ is the polytropic index. Sound waves travel at a speed
\beq\label{e.soundspeed}
c = \left[\left({n+1 \over n}\right) {\Pi \over \Sigma}\right]^{1/2},
\eeq
and a fluid element feels a
force\footnote{Throughout this paper ``force'' refers to force per unit
mass.} $-\nabla \Pi/\Sigma$ in the presence of a pressure
gradient. The force can also be written as the negative gradient
of the enthalpy $H$, where
\beq\label{e.enthalpy}
H=K(n+1)\Sigma^{1/n}.
\eeq
The gravity of the disk is ignored.

A disk of non-interacting particles (or pressure-free fluid elements)
on aligned Kepler orbits has a surface density given by \citep{Sta99}
\beq\label{e.surfacedensity}
\Sigma(a,E) = {\mu(a) \over 2 \pi a} {(1-e^2)^{1/2} \over 1 - e^2
        - a e^\prime (e + \cos E)}.
\eeq
In equation (\ref{e.surfacedensity}),
$a$ is the semimajor axis and $E$ is the eccentric anomaly. The
arbitrary function $\mu(a)$ gives the mass per unit interval of $a$,
and the eccentricity profile is described by $e(a)$ and its first derivative
$e^\prime(a)$. Equation (\ref{e.surfacedensity}) is valid only if the
orbits do not cross, which requires $|e + a e^\prime| < 1$; but $e$ is
{\em not\/} assumed to be small. (A more general version of this formula
for unaligned orbits is given in Appendix A.)

Clearly if $e^\prime=0$, then $\Sigma=\Sigma(a)$ and the density is
constant around each orbit.
This happens because the speed at each point around the orbit is inversely
proportional to the local separation between neighboring orbits,
and so competing terms in the equation of continuity cancel. Now,
treating the pressure as a perturbation,
the equation of state (\ref{e.eos}) implies that the pressure is stratified
on the unperturbed orbits. The perturbing force on each fluid element is
therefore perpendicular to the orbit, and does no work. Its
primary effect is to drive a precession, at a rate $\Omega_p$ which
must be constant at all radii for the disk to be long-lived at finite
eccentricity. This requirement is sufficient to determine the radial density
profile of the disk, as follows:
The small dimensionless parameter in the problem is $\epsilon =
\Omega_p/\omega$, where $\omega$ is the Keplerian mean motion.
Since the latter scales as $a^{-3/2}$, $\Omega_p={\rm constant}$ implies
$\epsilon \sim a^{3/2}$. To leading order, $\epsilon$ must be
proportional to the ratio of the perturbing force $-\nabla \Pi/\Sigma$
to the Keplerian force $GM_c/r^2$, where $M_c$ is the central
mass. These quantities should each be thought of as averaged over the
unperturbed orbit; since $e={\rm constant}$, these averages are
proportional to $-(d\Pi/da)/\Sigma$ and $GM_c/a^2$, respectively. The
required scaling for $\epsilon$ then implies that $(d\Pi/da)/\Sigma \sim
a^{-1/2}$, which can be written as
\beq
K \left(1 + {1 \over n}\right) \Sigma^{-1+1/n} d \Sigma
= - \left({a \over a_\ast}\right)^{-1/2} da,
\eeq
where $a_\ast$ is an arbitrary constant. Integrating this equation and
defining $a_1\equiv [K(n+1)/2(n-1)]^2 a_\ast$ gives the density profile
\beq\label{e.densityprofile}
\Sigma(a) = \Sigma_0 \left[1 - (a/a_1)^{1/2}\right]^n,
\eeq
where the surface density is equal to $\Sigma_0$ at the center and falls
to zero at $a=a_1$.

In the limit $n\to \infty$ the equation of state (\ref{e.eos}) becomes
isothermal and one obtains the density law
\beq
\Sigma = \Sigma_0 \exp\left[-(a/h)^{1/2}\right],
\eeq
where $h$ is a constant scale length. However, this disk is
infinite in extent, requiring $\epsilon$ to become arbitrarily large
since it must scale as $a^{3/2}$. I therefore exclude the isothermal case as
unphysical.

\subsection{Precession Rate and Surface Density\label{s.precession}}

Fluid elements in the disk move along perturbed Kepler orbits.
Ordinarily one would expect, in addition to the
secular precession, cyclic variations in the orbital elements
that would distort the pure ellipses and necessitate a first order
correction to the surface density. In this case, however, one can show that
the correction is zero.

Using equations (\ref{e.soundspeed}), (\ref{e.enthalpy}),
and (\ref{e.densityprofile}), the enthalpy can be written as
\beq\label{e.altenthalpy}
H=n c_0^2 \left[1 - (a/a_1)^{1/2}\right],
\eeq
where $c_0$ is the sound speed at $a=0$. The radial and tangential
components, ${\bar R}$ and ${\bar T}$, of the perturbing force are therefore
given by
\begin{eqnarray}\label{e.pressureforce}
({\bar R},{\bar T}) &=& -{dH \over da} \left({\partial a \over \partial r},
	{1 \over r} {\partial a \over \partial f} \right) \nonumber \\
&=& {n c_0^2 \over 2 (1-e^2) (a a_1)^{1/2}} (1+e \cos f, -e \sin f),
\end{eqnarray}
where $f$ is the true anomaly and $a=r(1+ e\cos f)/(1-e^2)$.
But the Keplerian velocity has components
\beq
(v_R, v_T) = {(GM_c)^{1/2} \over a^{1/2} (1-e^2)^{1/2}} (e \sin f, 1+e \cos f);
\eeq
consequently, in the frame that rotates at speed $\Omega_p$,
the Coriolis term $-2 \Omega_p {\hat z} \times \mbox{\boldmath $v$}$ will
identically cancel the pressure gradient if
\beq\label{e.precessionrate}
\Omega_p=-{n c_0^2 \over 4 [ G M_c a_1 (1-e^2)]^{1/2}}.
\eeq
In this case, the disk is globally in {\em geostrophic balance\/},
and the only force
term surviving in the Euler equation is the Kepler term
$-G M_c/r^2$. (The centrifugal term $\Omega_p^2 r$ is second order.) Thus
{\it in the rotating frame\/} the fluid elements follow exact Kepler
ellipses, and the surface density is given by equation
(\ref{e.densityprofile}) with no first-order correction.

It is easy to see intuitively why this result holds. Since the density is
constant around each orbit, the perturbing force is proportional to
the pressure gradient, which is inversely proportional to the separation
between isobars. But the isobars are the orbits, whose
separation is inversely proportional to the orbital speed when
$e = {\rm constant}$. Thus the perturbing
force is both proportional to the velocity in magnitude and perpendicular
to it in direction, just as the Coriolis force is. The two forces can cancel
on all orbits for the {\em same\/} $\Omega_p$ owing to the $a^{1/2}$
dependence of the enthalpy.

One can verify that equation (\ref{e.precessionrate}) agrees with the
precession rate obtained from standard perturbation theory by
using equation (\ref{e.pressureforce}) in the formula for the instantaneous
drift rate of the argument of pericenter,
\beq\label{e.burnsrate}
{\dot \varpi} = {1 \over e} \left[{a (1 - e^2) \over G M_c}\right]^{1/2}
\left( -{\bar R} \cos f + {\bar T} \sin f {2 + e \cos f
\over 1 + e \cos f} \right)
\eeq
\citep{Bur76,MD99}, and averaging around the orbit.
Equation (\ref{e.precessionrate}) may also be written in the form
\beq
\Omega_p=-{n \over 4 (1-e^2)^{1/2}} \left({c_0 \over v_1}\right)^2 \omega_1,
\eeq
where $v_1$ and $\omega_1$ are the Keplerian speed and mean motion at
the disk edge. This shows that the expansion parameter $\epsilon$ is basically
$a^{3/2}$ times
the square of the ratio of the central sound speed to the edge Keplerian speed.
Notice also that the precession is retrograde, and is faster
for larger eccentricities.

\begin{figure}[t]
{\epsfxsize=6in\hfil\epsfbox{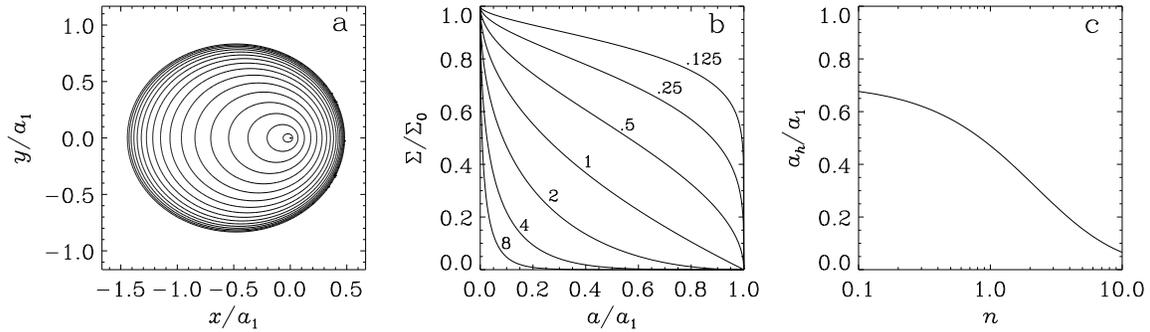}}
\caption{\footnotesize
($a$) Surface density contours for the disk with $n=3$ and $e=0.5$.
Contours are logarithmically spaced by factors of 2. ($b$) Surface density
as a function of semimajor axis for models with values of the polytropic index
$n$ as indicated. ($c$) Semimajor axis enclosing half the mass as a function
of $n$.
\label{f.density}}
\end{figure}

Figure \ref{f.density}a shows a density contour plot for a disk with $n=3$ and
$e=0.5$, illustrating the stratification of the density on orbits. Figure
\ref{f.density}b compares the radial density profiles
for selected values of the index $n$. The total mass of the disk is
\beq
M=24 \pi {\Gamma(n+1) \over \Gamma (n+5)}(1-e^2)^{1/2} \Sigma_0 a_1^2 ,
\eeq
where $\Gamma(x)$ denotes the gamma function.
The semimajor axis $a_h$ enclosing half the mass is given by
the solution of
\beq
I_z(4,n+1)={1 \over 2},
\eeq
where $I_x(p,q)$ is the incomplete beta function and
$z\equiv (a_h/ a_1)^{1/2}$. Figure \ref{f.density}c shows the variation
of $a_h$ with $n$. For $n=0$ the density is constant and
$a_h=2^{-1/2}a_1$; as $n$ increases the disks become increasingly
centrally concentrated.

\subsection{Behavior at the Center\label{s.center}}

The behavior of the models near their centers presents a problem. It
is not realistic to expect $e$ to remain finite as $a \to 0$, since there
are no strongly eccentric central masses in astrophysical systems. One could
imagine carving out an inner hole from the disk to avoid the center; but
this expedient would remove the pressure gradient needed to synchronize
the precession of the innermost material with that of the rest of
the disk. Instead, the eccentricity needs to go smoothly to zero at the
center---or at small finite radii---while maintaining the precession rate.
I show here how this transition may be effected by allowing the density to
deviate from that in equation (\ref{e.densityprofile}) over
a small interval of $a$, in the approximation where the density and
eccentricity changes are both small (in a sense defined below). Formally
this assumption limits the treatment to eccentricities $\lesssim 0.3$;
but it seems reasonable that a similar approach can be taken for larger $e$.

Readers more interested in the results than in the technical details are
invited to skip down to the last paragraph of this section.

At leading order, the changes to the precession frequency from the altered
density and eccentricity profiles  can be considered separately and summed.
I begin with the former, replacing the surface density $\Sigma$ by
$\Sigma [1 + \delta(a)]$, where $\delta$ is the fractional density increase.
The equation of state gives
\beq\label{e.deltagradient}
{\nabla \Pi \over \Sigma} = \left({\nabla \Pi \over \Sigma}\right)_{\delta=0}
+{\delta \over n} \left({\nabla \Pi \over \Sigma}\right)_{\delta=0}
+ K (1+n^{-1}) \Sigma_{\delta=0}^{1/n} \nabla \delta,
\eeq
where, from equations (\ref{e.enthalpy}) and (\ref{e.altenthalpy}),
$K=[n/(n+1)]c_0^2 \Sigma_0^{-1/n}$.
The first term on the right-hand side is responsible for the precession
given in equation (\ref{e.precessionrate}). Because $\delta$ is
constant on orbits, the second term simply contributes an additional
$\delta/n$ times the same rate. The third term is more difficult. The
radial and tangential components of $\nabla \delta$ are
\beq
{\partial\delta \over \partial r} =
\delta^\prime {1 + e \cos f \over 1-e^2}, \qquad
{1 \over r}{\partial\delta \over \partial f} =
-\delta^\prime  { e\sin f \over 1 - e^2},
\eeq
where $\delta^\prime \equiv d\delta/da$. Using equations
(\ref{e.densityprofile}), (\ref{e.deltagradient}), and the relations
\beq\label{e.ftoe}
\cos f = {\cos E - e \over 1 - e \cos E}, \qquad
\sin f = {(1 - e^2)^{1/2} \sin E \over 1 - e \cos E},
\eeq
this results in the force components
\beq
({\bar R}, {\bar T})
= c_0^2 \left[1 - (a/a_1)^{1/2} \right] \delta^\prime
\left[ -{1 \over 1 - e \cos E}, {\sin E \over
(1-e^2)^{1/2} (1 - e \cos E)} \right].
\eeq
The instantaneous precession
rate can be obtained from equation (\ref{e.burnsrate}), written in terms
of the eccentric anomaly using equation (\ref{e.ftoe}). Because
$\omega dt = (1- e \cos E) dE$, the resulting expression can be time-averaged
around the orbit by multiplying by $(1-e \cos E)$ and averaging over $E$.
(Useful formulae for evaluating the necessary averages are given
in Appendix B.) Adding the contributions of both the second and third
terms in equation (\ref{e.deltagradient}), the total extra precession induced
by the density enhancement $\delta$ is found to be
\begin{eqnarray}
\Omega_{p\delta} &=& -{c_0^2 \over 2 \left[GM_c a_1 (1-e^2)\right]^{1/2}} \nonumber \\
 & & \times 
\left\{ {\delta \over 2} - {\delta^\prime [(a a_1)^{1/2}-a] \over e}
\left[1 + 2 {(1-e^2)(1-e)\left(1-[1-e^2]^{1/2}\right)
\over e^2} \right]\right\}.
\end{eqnarray}
Note that $\Omega_{p\delta} < 0$ for positive density enhancements that
diminish outward.

To calculate the corresponding result for the eccentricity gradient,
$\Sigma$ is replaced by $\Sigma [1 + g(r,f)]$, where, from equation
(\ref{e.surfacedensity}),
\beq
g(r,f) = e^\prime r {2e + (1+e^2) \cos f \over (1-e^2)^2}.
\eeq
The equation of state gives, to leading order in $g$,
\beq\label{e.eprimegradient}
{\nabla \Pi \over \Sigma} = \left({\nabla \Pi \over \Sigma}\right)_{e^\prime=0}
+{g \over n} \left({\nabla \Pi \over \Sigma}\right)_{e^\prime=0}
+ K (1+n^{-1}) \Sigma_{e^\prime=0}^{1/n} \nabla g.
\eeq
As in equation (\ref{e.deltagradient}), the first term on the right hand
side gives the ordinary precession rate (\ref{e.precessionrate}).
However, the second term does not produce a proportional contribution
because $g$ is not constant around the orbit. Instead, it produces the force
components
\beq
({\bar R}, {\bar T}) = 
{c_0^2 \over 2} \left({a\over a_1}\right)^{1/2} { e^\prime \over 1-e^2}
\left[{\cos E + e \over 1 - e \cos E},
-{e \sin E (\cos E + e) \over (1-e^2)^{1/2}(1-e \cos E)} \right].
\eeq
The third term is algebraically cumbersome because of the gradient of $g(r,f)$.
Defining
\begin{mathletters}
\begin{eqnarray}
g_1(a) &=& e (e^\prime + a e^{\prime\prime}) + 2a {1 + e^2
	\over 1 - e^2}e^{\prime 2}, \\
g_2(a) &=& e^\prime + a e^{\prime\prime} + {4 a e e^{\prime 2} \over 1-e^2}, \\
g_3(a) &=& e g_1(a) + e^\prime, \\
g_4(a) &=& e [g_2(a) - e^\prime],
\end{eqnarray}
\end{mathletters}
the components of $\nabla g$ can be written as
\beq
{\partial g \over \partial r} = {g_1(a) + g_2(a) \cos E \over
	(1-e^2)(1 - e \cos E)}, \qquad
{1 \over r} {\partial g \over \partial f} = -{[g_3(a) +g_4(a) \cos E] \sin E
	\over (1 - e^2)^{3/2} (1 - e \cos E)}.
\eeq
This produces the force components
\beq
({\bar R}, {\bar T}) = 
c_0^2 \left[1 - (a/a_1)^{1/2} \right] { e^\prime \over 1-e^2}
\left\{-{g_1(a)+g_2(a)\cos E \over 1 - e \cos E},
{ [g_3(a)+g_4(a)\cos e]\sin E \over (1-e^2)^{1/2}(1-e \cos E)} \right\}.
\eeq
Following the same procedure as above, I find the total extra precession
induced by the eccentricity gradient to be
\beq
\Omega_{p e^\prime} = {c_0^2 \over (GM_c a_1)^{1/2} } {1 \over e (1-e^2)^{3/2}}
\left[-{ae^\prime \over 4}+{(aa_1)^{1/2}-a \over e^2}\sum_{i=1}^4 c_i(a) g_i(a)
\right],
\eeq
where the $c_i$ coefficients are given by
\begin{mathletters}
\begin{eqnarray}
c_1(a) &=& -e (1-e^2) \left[1 - (1-e^2)^{1/2} \right], \\
c_2(a) &=& (1-e^2)^{3/2} \left[1 - (1-e^2)^{1/2} \right], \\
c_3(a) &=& 1 - {e^2 \over 2} - (1-e^2)^{3/2}, \\
c_4(a) &=& {1-e^2 \over e} \left[1 - {e^2 \over 2} - (1-e^2)^{1/2} \right].
\end{eqnarray}
\end{mathletters}

The inner transition region, where $e$ is varying and $\delta$ is nonzero,
should join smoothly onto the main body of the disk, where
$e=e_0={\rm constant}$. The requirement that both regions precess
together is given by
\beq
\Omega_p + \Omega_{p\delta} + \Omega_{p e^\prime} = \Omega_{p0},
\eeq
where $\Omega_{p0}$ is the main body precession rate, i.e., equation
(\ref{e.precessionrate}) evaluated at $e=e_0$. This requirement yields a
differential equation linking $e(a)$ and $\delta(a)$:
\begin{eqnarray}\label{e.deltaequation}
\lefteqn{
{2 \delta^\prime \over e^3}[(aa_1)^{1/2}-a]
\left\{ e^2+2 (1-e^2)(1-e)[1-(1-e^2)^{1/2}]\right\} \nonumber
}
\\
&& =\delta +n \left[ \left({1-e^2 \over 1-e_0^2}\right)^{1/2} -1 \right]
+ {1 \over e(1-e^2)} \left[-a e^\prime + 4{(aa_1)^{1/2}-a  \over e^2}
\sum_{i=1}^4 c_i(a) g_i(a)\right].
\end{eqnarray}
Since the point of the exercise is to manipulate $e$, I
let $e(a)$ be a fixed function and 
use equation (\ref{e.deltaequation}) to calculate $\delta(a)$.
In order to avoid a discontinuity in the gradient of the surface density,
$e(a)$, $e^\prime(a)$, and $e^{\prime\prime}(a)$ should all be continuous
at $a=a_0$. A function with the desired properties is
\beq\label{e.centraleofa}
e(a) = \left\{ \begin{array}{l@{\quad,\quad}l}
	e_0 \left[ 1 + \left(1 - {a \over a_0}\right)^b
		\left(\sin{\pi a \over 2 a_0}-1 \right)\right] & a \leq a_0; \\
	e_0 & a_0 < a \leq a_1.
	\end{array} \right.
\eeq
For $0 < b < 1$, $e(a)$ is linear at small $a$ and joins smoothly onto
$e=e_0$ at $a=a_0$.

\begin{figure}[t]
{\epsfxsize=4.5in\hfil\epsfbox{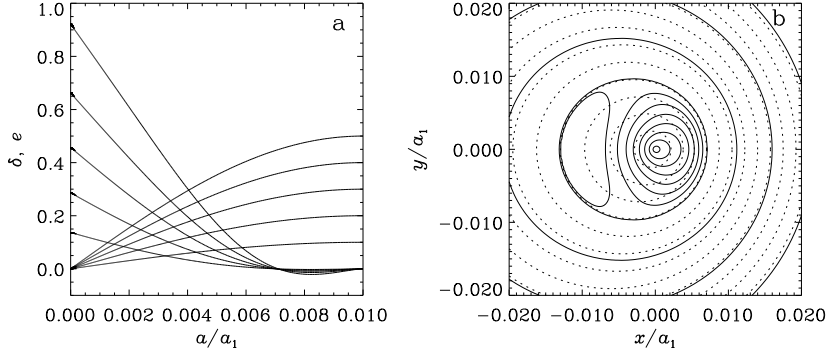}}
\caption{\footnotesize
($a$) Eccentricity $e$ ({\em thin lines\/}) and fractional density
enhancement $\delta$ ({\em thick lines\/}) in the small inner region where
$e$ goes to zero according to equation (\ref{e.centraleofa}), for models
with $n=3$. Successive curves, bottom to top,
show results for main-body ellipticities $e_0 = 0.1,0.2,0.3,0.4,0.5$.
($b$) Surface density contours ({\em solid lines\/}) in the inner region
of the model with $n=3$ and $e_0=0.3$. Contours are logarithmically spaced
by factors of $2^{1/8}$. {\em Dotted lines\/} show representative fluid
streamlines.
\label{f.center}}
\end{figure}

Figure \ref{f.center}a shows profiles of $e$ and the fractional density
enhancement $\delta$ in a transition region with radius $a_0=0.01a_1$ and
$b=1/2$, for a set of disks with $n=3$
and various eccentricities $e_0$. The results are fairly insensitive to
the value of $n$. One sees that the eccentricity can be made to go
continuously to zero at the center by steepening the density profile
through most of the transition region. Figure \ref{f.center}b shows
density contours in the transition region for the disk with $e_0=0.3$.
While the density is no longer constant along streamlines, the variation
about the mean is in the worst case only $\pm 20 \%$.
For $e_0 \gtrsim 0.3$, the needed density enhancement becomes
of order unity, which violates the assumption behind equation
(\ref{e.deltagradient}). The assumption that $g(r,f)$ is also small
requires that $ae^\prime \ll 1-e$, a condition that is consistent with,
but more stringent than, that for the non-crossing of orbits.
The choice of $e(a)$ in equation (\ref{e.centraleofa}) begins to violate
this condition for $e_0 \gtrsim 0.3$. But these conditions are required only
by the first-order expansion; a more careful treatment may be able
to find similar remedies for larger $e_0$.

\section{Connection with Axisymmetric Disks\label{s.axisymmetric}}

For small $e$, the structure of the models can also be understood in terms of
the acoustic modes of an axisymmetric disk. A formalism for dealing with
such disturbances is given by \citet{PaS91} and \citet[hereafter
HPS]{HPS92}. They start with a perturbation to the surface density of
the form
\beq\label{e.perturbation}
\Sigma=\Sigma(r) + \Sigma^\prime(r) e^{i(m\phi + \sigma t)},
\eeq
where $\Sigma(r)$ refers to the unperturbed disk; similar forms are assumed
for the other hydrodynamic quantities. They then merge the linearized fluid
equations into a single operator equation (HPS's eqs.\ [9] and [10]):
\begin{mathletters}
\beq
\Sigma^\prime = {\cal L}(W),
\eeq
where
\begin{eqnarray}
{\cal L}(W) &\equiv& -{1 \over r} {d \over dr} \left( {r \Sigma \over D}
\left[ {dW \over dr} + {2 m \Omega {\bar \sigma} W \over \kappa^2 r} \right]
\right) \nonumber\\
&& + 
\left[ {dW \over dr} + {2 m \Omega {\bar \sigma} W \over \kappa^2 r} \right]
{2 m \Omega {\bar \sigma} \Sigma \over \kappa^2 r D} + {m W \over {\bar
\sigma} r} {d \over dr} \left( {1\over \zeta}\right) - {4 m^2 \Omega^2
\Sigma W \over \kappa^4 r^2},\\
\kappa^2 &\equiv& {2 \Omega \over r} {d (r^2 \Omega) \over dr}, \\
\zeta &\equiv& {1 \over r \Sigma} {d (r^2 \Omega) \over dr},\\
{\bar \sigma} &\equiv& \sigma + m \Omega, \\
D &\equiv& {\bar \sigma}^2 - \kappa^2,
\end{eqnarray}\label{e.hpsequation}
\end{mathletters}
and $\Omega$ is the circular frequency.
In the simple case where the only gravity is that of the central mass,
$W=c^2 \Sigma^\prime / \Sigma$, and equation (\ref{e.hpsequation})
(rather than HPS's more complicated equation [15]) governs the modes.

In the axisymmetric limit, the present models have a surface density given by
\beq
\Sigma(r)=\Sigma_0 \left[1-(r/a_1)^{1/2}\right]^n
\eeq
and a sound speed given by
\beq
c^2(r) = c_0^2 \left[ 1- \left( r / a_1 \right)^{1/2}\right].
\eeq
The pressure gradient reduces $\Omega$ from its Keplerian value to
\beq
\Omega = \left({G M_c \over r^3}\right)^{1/2} -
{n c_0^2 \over 4 (G M_c a_1)^{1/2}}.
\eeq
The difference in surface density between the eccentric and axisymmetric
disks is, in the limit of small $e$,
\beq
\Delta \Sigma = \Sigma_0 \left\{ \left[ 1 - \left( r / a_1
\right)^{1/2} \left( 1 - {e \over 2} \cos f \right) \right]^n - \left[
1- \left( r / a_1 \right)^{1/2}\right]^n \right\}.
\eeq
For arbitrary $n$, $\Delta \Sigma$ is a superposition of Fourier
harmonics. I restrict the discussion here to the case $n=1$, so that
\beq
\Delta \Sigma = -{\Sigma_0 e \over 2} \left( {r \over a_1} \right)^{1/2}\cos f.
\eeq
This corresponds to a single $m=1$ term, with
\beq
\Sigma^\prime = {\Sigma_0 e \over 2} \left( {r \over a_1} \right)^{1/2},
\eeq
and consequently
\beq
W = {c_0^2 e \over 2} \left( {r \over a_1} \right)^{1/2}.
\eeq
Direct substitution into equation (\ref{e.hpsequation}b) shows, after
lengthy algebra, that equation (\ref{e.hpsequation}a) is satisfied to
leading order in $c_0^2$ when $\sigma = c_0^2/[4(GM_c a_1)^{1/2}]$. The
form of the exponential in equation (\ref{e.perturbation}) indicates
that this value corresponds to a {\em regression\/} of the pattern and
agrees with equation (\ref{e.precessionrate}) for $n=1$ and $e=0$. Also,
since $\sigma$ is real, the mode is stable.

The reader may have noticed that the
inner and outer radial boundary conditions have not played a role in
this discussion. The fact that $\Sigma^\prime$ is real means that the
mode has zero radial wavenumber. Unlike WKB waves, this wave is completely
unwound and propagates only in the tangential direction. In this respect
it is not really a ``mode'' in the sense of being a standing wave in a
resonant cavity. Instead, it is a traveling wave. In fact, since the
wave does not compress the fluid---even though the fluid itself is {\em not\/}
incompressible---it bears a close similarity to an ocean wave. Like an
ocean wave, fluid elements move epicyclically with respect to the mean
flow, and the restoring force for the pressure driven vertical displacement
is the gravity of the central mass.
The variation of the sound speed and flow velocity with depth
conspire to refract the wave around the central mass while keeping the
wave front straight.

\section{Line Profiles\label{s.lineprofiles}}

Constant-eccentricity Keplerian disks are kinematically simple. Spatially
resolved rotation curves along any line through the central mass will show
velocities proportional to $r^{-1/2}$, but will
not, in general, be antisymmetric about the center. The observed
velocities on opposite sides of the central mass will differ by a
multiplicative factor depending on both the eccentricity and the viewing
geometry. Kinematic measurements along several position angles would be
necessary to determine the disk structure and orientation; but a
reasonably accurate estimate of the central mass could be obtained by
averaging the masses derived from each side of the rotation curve under
the naive assumption of circular motion.

It seems more likely, however, that disks of the sort discussed here
will usually be unresolved. The expected emission line profiles from an
unresolved disk can be straightforwardly calculated.
For simplicity, I assume the emission comes from
a recombination line of a species whose density follows the total density,
so that the emissivity is proportional to $\Sigma^2$. The disk
is assumed to be optically thin with the line of sight in the disk
plane, an angle $\theta$ from the major axis.
A mass $dM$ on a single Kepler orbit in the disk produces a contribution
to the line of sight velocity distribution (LOSVD) $f(v_\ell)$ given by
\beq
df(v_\ell) = {dM \over 2 \pi (GM_c/a)^{1/2}}
{(1-e \cos E)^3 \over (\cos E -e) \cos \theta + (1-e^2)^{1/2}\sin E \sin\theta}.
\eeq
At each $v_\ell$, the contribution must be summed over the two
values of $E$ given by
\beq
\sin E_1 = {A - B (A^2+B^2-1)^{1/2} \over A^2 + B^2}, \qquad
\cos E_1 = {B + A (A^2+B^2-1)^{1/2} \over A^2 + B^2},
\eeq
and
\beq
\sin E_2 = {A + B (A^2+B^2-1)^{1/2} \over A^2 + B^2}, \qquad
\cos E_2 = {B - A (A^2+B^2-1)^{1/2} \over A^2 + B^2},
\eeq
where
\begin{mathletters}
\begin{eqnarray}
A &\equiv& -{(GM_c/a)^{1/2}\over v_\ell} \cos \theta,\\
B &\equiv& {(GM_c/a)^{1/2} \over v_\ell} (1-e^2)^{1/2} \sin \theta + e.
\end{eqnarray}
\end{mathletters}
These expressions ignore the precession of the disk, which adds a slow,
retrograde, solid-body component to the rotation; this effect is included
in the examples shown below, but makes only a tiny difference to the results.
Assuming an ideal gas, the equation of state and the density profile
imply a one-dimensional thermal velocity dispersion
\beq
\sigma_{\rm th}^2 = {n c_0^2 \over n+1} \left[ 1 - (a/a_1)^{1/2} \right].
\eeq
The LOSVD at each $a$ is convolved to the thermal dispersion, and the results
are integrated over the disk. No correction is made for a transition to
$e=0$ at the center.

\begin{figure}[t]
{\epsfxsize=4in\hfil\epsfbox{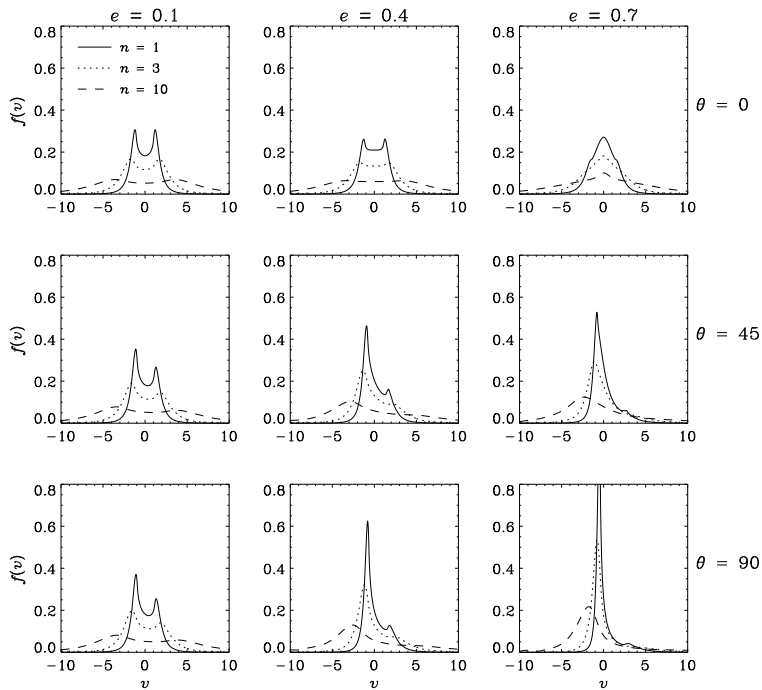}}
\caption{\footnotesize
Simulated line profiles for unresolved elliptic disks. Relative
fluxes per unit interval of velocity are shown for velocities in units
of $(GM_c/a_1)^{1/2}$. Results for eccentricities of $0.1$, $0.4$, and
$0.7$ are shown in the first, second and third columns, and for lines
of sight along, oblique to, and perpendicular to the disk major axis
in the top, middle, and bottom rows. Different line styles
correspond to values of the polytropic index $n$ as indicated at top left.
The central sound speed is $0.2$.
\label{f.lineprofiles}}
\end{figure}

\begin{figure}[t]
{\epsfxsize=6.4in\hfil\epsfbox{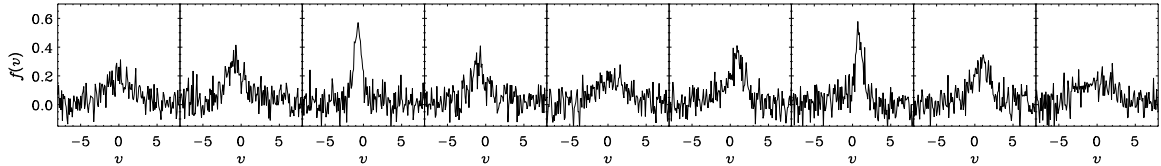}}
\caption{\footnotesize
A sequence of line profiles for the $n=3$, $e=0.7$ model, shown
({\em left to right\/}) as the disk precesses through a complete period.
Gaussian noise has been added to simulate observations. The true mean velocity
is zero, but continuum misidentification could lead to spurious measurements
of velocity offsets.
\label{f.voffset}}
\end{figure}

Line profiles for various models are shown in Figure \ref{f.lineprofiles}.
The line shape appears at first glance very sensitive to the polytropic index
$n$. But much of this sensitivity is a simple scaling; for the same
central mass and limiting radius, disks of larger $n$ show broader line
profiles simply because they are more centrally concentrated. One also
sees that the lines can be strongly asymmetric, even for eccentricities as
small as $0.1$. Where the profiles are double peaked, the stronger peak
arises from the side of the disk containing the apocenters of the orbits,
and consequently can appear at either positive or negative velocity.
For eccentricities of $0.5$ or larger, the secondary peak can disappear
entirely. Even though the line centroid remains at zero velocity
relative to the central mass, in the presence of noise the asymmetric
primary peak could easily be mistaken for a single narrow line with
a large velocity offset. Figure \ref{f.voffset} shows a sequence of line
profiles through a complete precession period for the $n=3$, $e=0.7$
disk, with noise added. In principle, the slow time evolution of the line
profile could be detectable for some types of objects. For a cold
protostellar disk, equation (\ref{e.precessionrate}) implies a precession
period of
\beq
T_p = 3900\, C \left( M_c \over 1 \msun \right)^{1/2}
	\left( a_1 \over 100 \au \right)^{1/2}
	\left( c_0 \over 1 \kms \right)^{-2} \yr,
\eeq
where the factor $C \equiv (4/n)(1-e^2)^{1/2}$ is of order unity.
For a galactic nuclear disk,
\beq
T_p = 4 \times 10^7\, C \left( M_c \over 10^8 \msun \right)^{1/2}
	\left( a_1 \over 1 \pc \right)^{1/2}
	\left( c_0 \over 10 \kms \right)^{-2} \yr.
\eeq

Offset or asymmetric emission lines are not uncommon in active galactic
nuclei (AGN), but are usually qualitatively different from those shown
above.  Systematic offsets, e.g., of the [\ion{O}{3}] line \citep{Whi85},
are consistently toward the blue, and it is the line centroid, rather
than the peak, that is shifted. Offsets of this sort should more naturally
arise from outflows rather than orbital motion. AGN showing double-peaked
lines characteristic of unresolved Keplerian motion may be modeled by
axisymmetric relativistic disks when the blue peak is stronger than the
red, the asymmetry arising from Doppler boosting \citep{CHF89}. However,
in a fraction of radio-loud AGN with disk-like H$\alpha$ emission, the red
peak is stronger than the blue. \citet{Era95} appeal to eccentric disks to
model these objects. Their models, like those presented here, have constant
eccentricity, but the line emissivity is taken to be a function of $r$
rather than $a$. \citet{Era95} consider general relativistic effects and
estimate that differential relativistic precession would ruin the coherence
of elliptic disks in AGN on timescales as short as decades, unless they
were narrow rings. \citet{Bao96} show that narrow rings can persist even in a
strong field, and calculate the resulting line profiles. But
it seems plausible that relativistic versions of the models of this paper could
be found in which combined relativistic and hydrodynamic effects could drive
a uniform precession, extending the lives of wide disks.

\section{Discussion\label{s.discussion}}

I have presented the properties of a particularly simple family of
idealized models
for eccentric Keplerian fluid disks, in which the internal pressure gradient
drives a coherent precession of the apses of the constant-eccentricity
streamlines. In the frame that rotates with the precession frequency, the
pressure gradient balances the Coriolis force; the disks are therefore in
global geostrophic balance, and the fluid motion is Keplerian, to first order
in the pressure, in the rotating frame.

In the limit of small eccentricity, the models can be viewed as perturbed
axisymmetric disks, and their elliptic distortions interpreted as traveling
acoustic waves. These waves are related to the ``slow'' $m=1$ modes
examined by \citet{Tre01}, but are not physically the same. The elliptic
distortions have radial wavenumber $k=0$, which is maintained by a continuous
refraction of the waves around the center, and therefore owe their existence
to the particular density profile of the disk. The disk's self gravity,
which creates the true standing waves studied by Tremaine, is ignored in
the present case. Including finite disk gravity may restrict
the slow acoustic waves to a limited range of radii and/or turn them into
standing waves with $k\neq 0$.

One should keep in mind that these models are highly idealized,
and neglect at least two important effects present in real disks. First,
fluid elements in a three-dimensional disk will experience a varying
vertical compression around every orbit, since the vertical ($z$) component
of the Kepler force near the disk plane is proportional to $z/r^3$. The
fractional magnitude of the variation depends on the eccentricity and not on
the disk thickness. Second,
shear viscosity is known to be a controlling factor in true
accretion disks, and is likely to affect the structure of larger-scale disks
in galactic nuclei which may not be actively accreting.

Turbulent viscosity
can arise from any of a number of local instabilities, the leading candidate
being the magnetorotational instability \citep{BaH91} when magnetic fields are
present. Angular momentum is transported efficiently by magnetic torques,
and the instability's growth time is of order the orbital period.
The instability will therefore affect the structure of any disk
with even a weak magnetic field, as long as the ionization is sufficiently
high that the flux is effectively frozen in the fluid. On the other hand,
the instability may be suppressed in protostellar \citep{War99} or
protoplanetary \citep{Rey01} disks of very low ionization.

A purely hydrodynamical instability unique to non-axisymmetric
disks has been studied by Goodman and collaborators \citep{Goo93,RyG94,RGV96}.
The ``eccentric instability''
is intimately connected with the inertial oscillations, force-free
epicyclic motions supported by any rotating or shearing medium.
Because the spectrum of the inertial oscillations is continuous, there is
always a mode of the right frequency available to be parametrically
amplified by a periodic disturbance, such as a traveling acoustic wave. The
instability is intrinsically three-dimensional---the motions of the
amplified mode are inclined to the disk plane---and therefore would
affect any finite-thickness analogue of the present models.

Turbulence may also be associated with 
vortices, whose importance in disks is considered by \citet{AdW95}. They point
out that, according to Kelvin's circulation theorem, vorticity should not be
spontaneously created when pressure and density contours coincide and
the viscosity is zero. Thus the models, as defined here, should not be
unstable to vortex generation. Real disks, however, may be radiatively
heated by their central objects and radiatively cooled by line or
continuum emission. The temperature could then vary significantly around
streamlines, in which case vortices could be generated. A careful
treatment of the radiative transfer within the models would be needed to
determine if this is the case.

\begin{figure}[t]
{\epsfxsize=2.5in\hfil\epsfbox{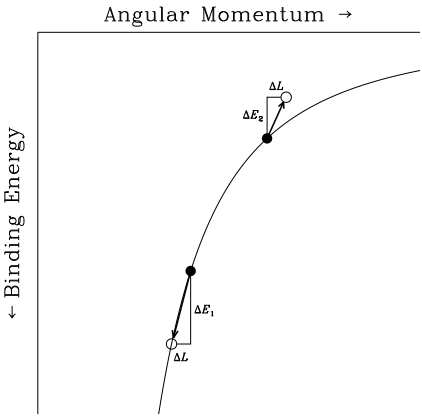}}
\caption{\footnotesize
Illustrating why a circular disk is not the minimum energy
configuration for a given angular momentum. A pair of circular
rings ({\em filled circles\/}) is plotted in the Lindblad diagram for a
Kepler potential. {\em Smooth curve\/} shows the circular orbits. {\em
Open circles\/} show the rings displaced so as to conserve total angular
momentum and lower the total energy ($|\Delta E_2| < |\Delta E_1|$),
leaving the outer ring eccentric.
\label{f.lindblad}}
\end{figure}

It may be tempting to argue that viscous dissipation will naturally circularize
any disk on the grounds that, at a fixed angular momentum, the circular
configuration is the state of lowest energy. However, this latter statement
is true only for single orbits, and does not apply to disks where
energy and angular momentum can be traded between orbits. To illustrate,
consider the toy problem of two circular rings of equal mass. Figure
\ref{f.lindblad} shows a Lindblad diagram for this system, with both rings
(filled circles) lying on the locus of circular orbits (smooth curve) for
the Kepler potential. Now imagine shifting the inner ring to a more tightly
bound circular orbit (lower open circle), lowering its energy by $\Delta E_1$
and its angular momentum by $\Delta L$. Angular momentum can be conserved
by depositing the same $\Delta L$ in the outer ring, along with some
amount of energy $\Delta E_2$. Because of the shape of the circular-orbit
locus, however, the shifted outer ring (upper open circle) can move away
from the circular orbits (i.e., can become eccentric) even if $|\Delta E_2| <
|\Delta E_1|$. {\em Thus, for any circular disk, there is an eccentric disk
with the same angular momentum and lower energy.}

Energy arguments alone are therefore not sufficient to determine whether
eccentric disks can survive in the presence of shear viscosity. A number
of more detailed calculations have been attempted, but
the results are still ambiguous. An analytic and numerical study
of the viscous evolution of pressure-free eccentric gas disks and streams
by \citet{SyC92} finds that the eccentricity nearly always {\em increases}
or remains constant under the influence of viscous effects. This result
is corroborated by the purely analytic work of \citet{LPP94}, who show
that evolution at constant $e$ is one of many possible paths for the
disk. \citet{Ogi01} finds, in contrast, that with gas pressure included,
eccentric motions dissipate on a viscous time scale. In three-dimensional
smooth-particle hydrodynamic simulations \citep{MaM98}, broad, persistently
eccentric disks are found to form spontaneously in accreting winds.  A more
careful analysis of the simulations would be needed to clarify the detailed
structure of the disks and the role of pressure in maintaining them.
Disks whose radial profiles are determined by accretion processes may
not evolve toward the special profiles required by the models in this
paper, and consequently may be unable to synchronize their
precession rates and remain eccentric. But the
appearance of long-lived eccentric disks in simulations where no special
effort is made to create them suggests that Nature may have
ways to make similar disks in abundance.

\acknowledgments

The author is grateful for helpful insights and suggestions from Mike
Eracleous, Glen Stewart, Joe Shields, Steve Balbus, and Larry Wilen.
This work was supported by NSF CAREER grant AST-9703036.

\section*{Appendix A. Surface Density of an Unaligned Keplerian Disk}

Equation (\ref{e.surfacedensity}) can be generalized to an unaligned
planar disk in which the eccentricity and longitude of pericenter are given
by functions $e(a)$ and $\varpi(a)$ respectively. The result is
\beq
\Sigma(a,E) = {\mu(a) \over 2 \pi a} {(1-e^2)^{1/2} \over 1 - e^2
-a \left[ e^\prime (e + \cos E) + \varpi^\prime e (1-e^2)^{1/2}\sin E \right]},
\eeq
where the eccentric anomaly $E$ is measured from pericenter on each orbit.
Alternatively, one might prefer to have the surface density in terms of
a fixed polar coordinate system $(r,\phi)$. In this case,
\beq
\Sigma(r,\phi) = {\mu(a) \over 2 \pi a}
{(1-e^2)^{1/2} \over 1 - e^2 - e^\prime [2ae +r\cos(\phi-\varpi)]
-r e \varpi^\prime \sin(\phi-\varpi)},
\eeq
where $a(r,\phi)$ can be found efficiently by iteration using the prescription
\beq
a_{j+1} = (1-k) a_j + k r {1 + re(a_j)\cos[\phi-\varpi(a_j)] \over
1 - [e(a_j)]^2}.
\eeq
Values of the relaxation factor $k\approx 0.5$ usually give rapid convergence.
The above results generalize the formulae of \citet{BGT86} to arbitrary
eccentricity, but remain valid only when the criterion for non-crossing of
orbits \citep{Ogi01},
\beq
(e + ae^\prime)^2 + (ae\varpi^\prime)^2 < 1,
\eeq
is satisfied.

\section*{Appendix B. Helpful Integrals}

The following integrals are useful in evaluating the average precession rates
in Section \ref{s.center}:
\begin{eqnarray}
{1 \over \pi} \int_0^\pi {dE \over 1 - e \cos E}  &=&
	{1 \over (1 - e^2)^{1/2}}, \\
{1 \over \pi} \int_0^\pi {\cos E dE \over 1 - e \cos E}  &=&
	{1 - (1 - e^2)^{1/2} \over e (1 - e^2)^{1/2}}, \\
{1 \over \pi} \int_0^\pi {\cos^2 E dE \over 1 - e \cos E} &=&
	{1 - (1 - e^2)^{1/2} \over e^2 (1 - e^2)^{1/2}}, \\
{1 \over \pi} \int_0^\pi {\sin^2 E dE \over 1 - e \cos E} &=&
	{1 - (1 - e^2)^{1/2} \over e^2}, \\
{1 \over \pi} \int_0^\pi {\sin^2 E \cos E dE \over 1 - e \cos E}  &=&
	{2-e^2-2(1-e^2)^{1/2} \over 2 e^3}, \\
{1 \over \pi} \int_0^\pi {\sin^2 E \cos^2 E dE \over 1 - e \cos E} &=&
	{2-e^2-2(1-e^2)^{1/2} \over 2 e^4}.
\end{eqnarray}


\clearpage
\begin{thebibliography}{}
\bibitem[Adams \& Watkins(1995)]{AdW95} Adams, F. C. \& Watkins, R. 1995,
	\apj, 451, 314
\bibitem[Bacon et al.(2001)]{Bac01} Bacon, R., Emsellem, E., Combes, F.,
	Copin, Y., Monnet, G., \& Martin, P. 2001, \aap, 371, 409
\bibitem[Balbus \& Hawley(1991)]{BaH91} Balbus, S. \& Hawley, J. 1991,
	\apj, 376, 214
\bibitem[Bao et al.(1996)]{Bao96} Bao, G., Hadrava, P., \& \O stgaard, E.
	1996, \apj, 464, 684
\bibitem[Borderies, Goldreich, \& Tremaine(1986)]{BGT86} Borderies, N.,
	Goldreich, P., \& Tremaine, S. 1986, Icarus, 68, 522.
\bibitem[Burns(1976)]{Bur76} Burns, J. A. 1976, Am. J. Phys., 44, 944
\bibitem[Chen, Halpern \& Filippenko(1989)]{CHF89} Chen, K., Halpern, J.
	P., \& Filippenko, A. V. 1989, \apj, 339, 742
\bibitem[Eracleous et al.(1995)]{Era95} Eracleous, M., Livio, M., Halpern,
	J. P., \& Storchi-Bergmann, T. 1995, \apj, 438, 610
\bibitem[Goodman(1993)]{Goo93} Goodman, J. 1993, \apj, 406, 596
\bibitem[Heemskerk(1994)]{Hee94} Heemskerk, M. H. M. 1994, \aap, 288, 807
\bibitem[Heemskerk, Papaloizou, \& Savonije(1992)]{HPS92} Heemskerk, M.
	H. M., Papaloizou, J. C., \& Savonije, G. J. 1992, \aap, 260, 161 (HPS)
\bibitem[Kormendy \& Bender(1999)]{KoB99} Kormendy, J., \& Bender, R.
	1999, \apj, 522, 772
\bibitem[Lauer et al.(1993)]{Lau93} Lauer, T. R., Faber, S. M., Groth, E. J.,
	Shaya, E. J., Campbell, B., Code, A., Currie, D. G., Baum, W. A.,
	Ewald, S. P., Hester, J. J., Holtzman, J. A., Kristian, J., Light, R.
	M., \& Westphal, J. A. 1993, \aj, 106, 1436
\bibitem[Lubow(1991)]{Lub91} Lubow, S. 1991, \apj, 381, 259
\bibitem[Lyubarskij, Postnov, \& Prokhorov(1994)]{LPP94} Lyubarskij, Yu. E.,
	Postnov, K. A., \& Prokhorov, M. E. 1994, \mnras, 266, 583
\bibitem[Mastrodemos \& Morris(1998)]{MaM98} Mastrodemos, N. \& Morris, M.
	1998, \apj, 497, 303
\bibitem[Murray(2000)]{Mur00} Murray, J. R. 2000, \mnras, 314, L1
\bibitem[Murray \& Dermott(1999)]{MD99} Murray, C. D. \& Dermott, S. F.
	1999, Solar System Dynamics, (Cambridge Univ. Press)
\bibitem[Ogilvie(2001)]{Ogi01} Ogilvie, G. I. 2001, \mnras, 325, 231
\bibitem[Osaki(1985)]{Osa85} Osaki, Y. 1985, \aap, 144, 369
\bibitem[Papaloizou \& Savonije(1991)]{PaS91} Papaloizou, J. C. \&
	Savonije, G. J. 1991, \mnras, 248, 353
\bibitem[Patterson(1998)]{Pat98} Patterson, J. 1998, \pasp, 110, 1132
\bibitem[Reyes-Ruiz(2001)]{Rey01} Reyes-Ruiz, M. 2001, \apj, 547, 465
\bibitem[Ryu \& Goodman(1994)]{RyG94} Ryu, D. \& Goodman, J. 1994, \apj,
	422, 269
\bibitem[Ryu, Goodman, \& Vishniac(1996)]{RGV96} Ryu, D., Goodman, J., \&
	Vishniac, E. T. 1996, \apj, 461, 805
\bibitem[Salow \& Statler(2001)]{Sal01} Salow, R. M. \& Statler, T. S.
	2001, \apjl, 551, L49
\bibitem[Statler(1999)]{Sta99} Statler, T. S. 1999, \apjl, 524, L87
\bibitem[Stehle(1999)]{Ste99} Stehle, R. 1999, \mnras, 304, 687
\bibitem[Syer \& Clarke(1992)]{SyC92} Syer, D. \& Clarke, C. J. 1992,
	\mnras, 255, 92
\bibitem[Tremaine(1995)]{Tre95} Tremaine, S. 1995, \aj, 110, 628 
\bibitem[Tremaine(2001)]{Tre01} Tremaine, S. D. 2001, \aj, 121, 1776
\bibitem[Vogt(1982)]{Vog82} Vogt, N. 1982 \apj, 252, 653
\bibitem[Wardle(1999)]{War99} Wardle, M. 1999, \mnras, 307, 849
\bibitem[Whitehurst(1988)]{Whi88} Whitehurst, R. 1988, \mnras, 232, 35
\bibitem[Whittle(1985)]{Whi85} Whittle, M. 1985, \mnras, 213, 1
\end{thebibliography}
\end{document}